# Molecule Oxygen Induced Ferromagnetism and Half-metallicity in $\alpha$-BaNaO$_4$: A First Principles Study

Jun Deng[†,‡], Jiangang Guo[†,§*], Xiaolong Chen[†,‡,§*]

[†]Beijing National Laboratory for Condensed Matter Physics, Institute of Physics, Chinese Academy of Sciences, Beijing 100190, China.

[‡]School of Physical Sciences, University of Chinese Academy of Sciences, Beijing 101408, China.

[§]Songshan Lake Materials Laboratory, Dongguan, Guangdong 523808, China.

**ABSTRACT:** Molecule oxygen resembles 3$d$ and 4$f$ metals in exhibiting long-range spin ordering and electron strong correlated behaviors in compounds. The ferromagnetic spin ordering and half-metallicity, however, are quite elusive and have not been well acknowledged. In this article, we address this issue to study how spins will interact each other if the oxygen dimers are arranged in a different way from that in the known super- and per-oxides by first principles calculations. Based on the results of structure search, thermodynamic study and lattice dynamics, we show that tetragonal $\alpha$-BaNaO$_4$ is a stable half-metal with a Curie temperature at 120 K, a first example in this class of compounds. Like 3$d$ and 4$f$ metals, the O$_2$ dimer carries a local magnetic moment 0.5 $\mu_B$ due to the unpaired electrons in its $\pi^*$ orbitals. This compound can be regarded as forming from the O$_2$ dimer layers stacking in a head to head way. Different from that in $A$O$_2$ ($A$=K, Rb, Cs), the spins are both ferromagnetically coupled within and between the layers. Spin polarization occurs in $\pi^*$ orbitals with spin-up electrons fully occupying the valence band and spin-down electrons partially the conduction band, forming the semiconducting and metallic channels, respectively. Our results highlight the importance of geometric arrangement of O$_2$ dimers in inducing ferromagnetism and other novel properties in O$_2$ dimer containing compounds.

## ■ INTRODUCTION

Molecule oxygen has long been known to carry a local moment because of the existence of unpaired electron in its $\pi^*$ orbitals.[1] When condensing into solid at low temperatures, ~24K, it exhibits a long-range antiferromagnetic (AFM) ordering.[2,3] Alkali metal superoxides (O$_2^-$), peroxides (O$_2^{2-}$) and dioxygenyl (O$_2^+$) salts all contain molecule oxygen but with different number of electrons in their $\pi^*$ orbitals, leading to a variety of magnetic properties. Typical examples are AFM $A$O$_2$($A$=Na, K, Rb, Cs),[4,5,6,7,8,9] spin-glass Rb$_4$O$_6$/Cs$_4$O$_6$,[10,11] ferrimagnetic (FiM) O$_2$PtF$_6$[12,13] and paramagnetic O$_2$SbF$_6$.[13] The long-range ferromagnetic (FM) ordering, however, is quite elusive in these O$_2$ dimer containing compounds though many attempts have been made over last decades.

A weak FM was reported in Ba$_{1-x}$K$_x$O$_2$ for $x$=0.269 and a spin-glass behavior for other K doping levels, which were ascribed to the coexistence and the competition between AFM and FM.[14,15] Partial oxygen deficiency at $x$=0.28 in RbO$_{2-x}$ was found to result in a transition from AFM to spin glass, and presumably there are short-range FM ordering in clusters at ~50 K.[16] Theoretically, the electrons in the $\pi^*$ orbitals of O$_2$ dimers can be better treated as a strong correlated system.[17,18] Naghavi $et\ al.$ predicted that exertion of pressure would make Rb$_4$O$_6$ transform from AFM insulator to FM insulator then to FM half-metal, which is a consequence of Mott transition.[19] It is argued to emerge aniono-genic ferromagnetism and half-metallicity in the interface of KO$_2$/BaO$_2$ through double exchange interactions of O$_2^{2-}$ and O$_2^-$ from the first principles calculations.[20] A question arises, is a long-range FM ordering possible in a bulky molecule oxygen containing compound as in a 3$d$ metal counterpart? If so, can a half metal be evolved from the FM? This is of importance in further understanding the interaction behavior among these local moments carried by O$_2$ dimers.

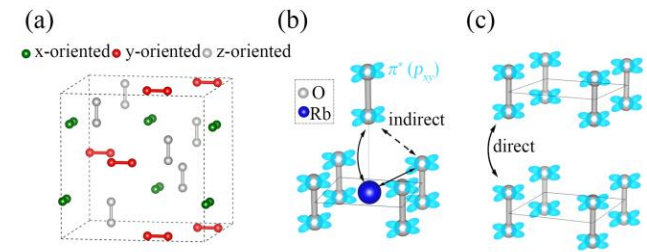

**Figure 1**. (a) Sketch of arrangement of O$_2$ dimers in Rb$_4$O$_6$. For simplification, all the cations are omitted. (b) Orbital based schematic diagram of indirect magnetic interaction for the O$_2$ dimers in RbO$_2$ and (c) hypothetically direct magnetic interaction of O$_2$ dimers. The magnetic interactions come from the unfilled $\pi^*$ ($p_{xy}$) orbitals (blue).

To begin with, we first show the importance of O$_2$ dimer orientation in inducing the long-range ordering by taking Rb$_4$O$_6$ and RbO$_2$ as an example, see Figure 1(a) and 1(b). The O$_2$ dimers in Rb$_4$O$_6$ align in three principle axes. Due to the geometric frustration, it does not show long-range magnetic orderings.[10] In contrast, the ground state of RbO$_2$ is experimentally confirmed to be AFM,[5] for all the O$_2$ dimers align parallel. The O$_2$ dimers couple ferromagnetically in the (001) planes in RbO$_2$ while interact antiferromagnetically between adjacent planes with a relative shift of (0.5a, 0.5a, 0), where a is the lattice parameter. Both magnetic exchange interactions are theoretically found to be indirect, i.e. via the cations through superexchnage.[21] So do for AFM KO$_2$ and CsO$_2$.[4] These facts suggest that the O$_2$ dimer layers in such a stacking way favor the AFM interlayer exchange interactions for local moments. Recent studies showed that the relative orientation of layers will give rise to tremendous changes in electronic properties and hence result in emergent phenomena.[22] We try to apply this idea to the O$_2$ dimer containing compounds to explore how their magnetic properties will change. Specifically, rearrange the relative orientation of neighboring O$_2$ dimer layers while keep the intralayer unchanged. Figure 1(c) shows a possible simple solution, where the O$_2$ dimers stack along the [001] direction in a head-to-head way. In this way, the O$_2$ dimers in neighboring planes are expected to have direct exchange interactions and bring about very different magnetic properties.

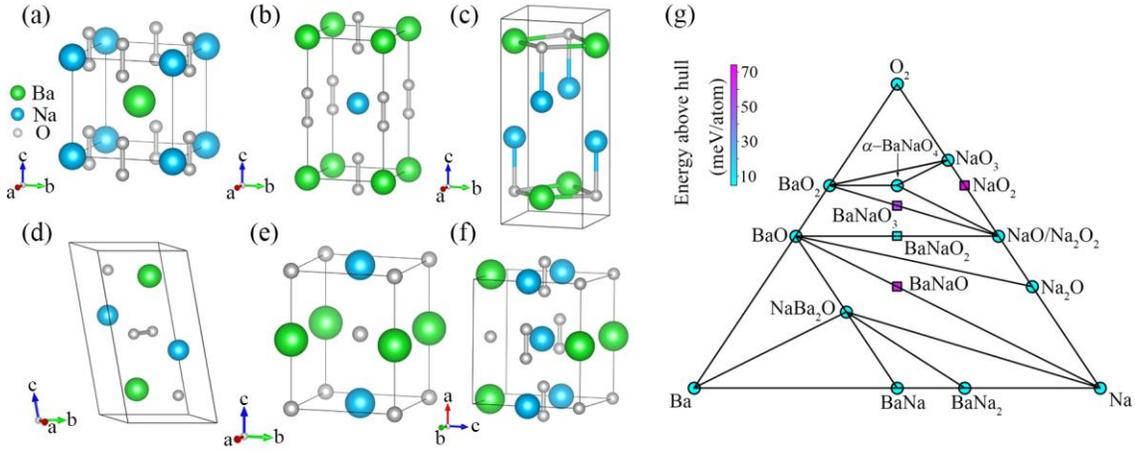

**Figure 2**. Crystal structures of (a) $P4/mmm$-BaNaO$_4$ ($\alpha$-BaNaO$_4$), (b) $P4/mmm$-BaNaO$_4$ ($\beta$-BaNaO$_4$), (c) $P4/nmm$-BaNaO, (d) $P\bar{1}$-BaNaO$_2$, (e) $P4/mmm$-BaNaO$_2$, and (f) $Pmmn$-BaNaO$_3$. Green balls stand for barium, blue ones for sodium and grey dumb bells for O$_2$ dimer. (g) Phase diagram based on convex hull analysis of formation enthalpies for the Ba-Na-O system at $T=0$ K. Formation enthalpy $\Delta H$ was defined as $\Delta H_{Ba_xNa_yO_z} = E_{Ba_xNa_yO_z} - xE_{Ba} - yE_{Na} - (z/2)E_{O_2}$, where $E_{Ba_xNa_yO_z}$, $E_{Ba}$, $E_{Na}$, $E_{O_2}$ is the total energy of Ba$_x$Na$_y$O$_z$, bcc-Ba, bcc-Na, and $\alpha$-O$_2$. The cyan circles indicate stable structures, while colored squares meta-stable/unstable.

Since no known binary peroxides and superoxides have such an O$_2$ arrangements in their structures, ternary compounds should be considered. However, very few ternary peroxides or superoxides have been reported so far. The substitution of K for Ba in BaO$_2$ only leads to a solid solution, without changing the structure type.[14] This is not beyond expectation for K$^+$ ion and Ba$^{2+}$ ion are comparable in size. It is more likely to form the desired structure if a smaller ion, say Na$^+$, is used instead.

In this Article, we report to identify $\alpha$-BaNaO$_4$ in which the O$_2$ dimers arrange in a way as shown in Figure 1(c) among the predicted compounds by systematic structure search. This compound is stable in terms of formation enthalpy, mechanical dynamics and lattice dynamics. It exhibits ferromagnetism and half-metallicity with a large spin gap ~4 eV (HSE06) in the ground state. The emergent half-metallicity is attributed to the partial occupation of the O$_2$ dimer $\pi^*$ bonds, which leads to the spin splitting. Monte Carlo simulations based on the Heisenberg model indicate that the spin ordering occurs at Curie temperature ($T_c$) of 120 K. Our results show that compounds with adjacent O$_2$ dimer layers stacking in a head-to-head way is possible and may change the interlayer coupling from AFM to FM, highlighting the importance of relative shift of adjacent layers in inducing novel properties for this category of compounds.

## COMPUTATIONAL DETAILS

The first principles calculations were carried out with the density functional theory (DFT) implemented in the Vienna *ab initio* simulation package (VASP).[23] We adopted the generalized gradient approximation (GGA) in the form of the Perdew-Burke-Ernzerhof (PBE)[24] for the exchange-correlation potentials. The projector augmented-wave (PAW)[25] pseudopotentials were used with a plane wave energy 680 eV. $2s^22p^4$, $2p^63s^1$, and $5s^25p^66s^2$ were treated as valence electrons for O, Na, and Ba, respectively. A Monkhost-Pack[26] Brillouin zone sampling grid with a resolution $0.02\times 2\pi$ Å$^{-1}$ was applied. Atomic positions and lattice parameters were relaxed until all the forces on the ions were less than $10^{-4}$ eV/Å. DFT+$U$ method was used where we additionally state. The Hubbard $U$ is applied to the O-$p$ orbitals. Phonon spectra were calculated using finite displacement method implemented in the PHONOPY code[27] to determine the lattice-dynamical stability of the structures. *Ab initio* molecular dynamic (AIMD) simulations were performed implemented in VASP. The time step was chosen as 3 fs, and the duration lasts 9 ps. The hybrid functional HSE06 with a mixing parameter of 25% for exact-exchange term was used to estimate the band gap in half metals. We used the code from Henkenlman Group[28] for Bader charge analysis. The structural searching and prediction were performed by the Particle Swarm Optimization (PSO) technique implemented in the CALYPSO code[29] in the Ba-Na-O system with molar ratio Ba:Na=1:1.

## RESULTS AND DISCUSSION

Structural search was performed on BaNaO$_x$ ($x$=1~4). The corresponding structures with low energy and reasonable coordination geometry were singled out to be as candidates for further investigations. This way yields *P*4/*mmm* α-BaNaO$_4$, *P*4/*mmm* β-BaNaO$_4$, *P*4/*nmm*-BaNaO, *P*$\bar{1}$-BaNaO$_2$, *P*4/*mmm*-BaNaO$_2$, and *P*mmn-BaNaO$_3$, see Figure 2(a-f). We note *P*4/*nmm*-BaNaO and *P*4/*mmm*-BaNaO$_2$ contain only O$^{2-}$; *P*$\bar{1}$-BaNaO$_2$ and BaNaO$_3$ both O$_2$ dimers and O$^{2-}$; α- and β-BaNaO$_4$ only O$_2$ dimers. β-BaNaO$_4$ resembles BaO$_2$ in structure except for the Ba at the body center being substituted by Na, which is added for comparison. α-BaNaO$_4$ can be regarded as an O$_2$ dimer deficient perovskite with a zero occupation of the dimer in the Ba layer. The O$_2$ dimers form a layer by bonding with Na$^+$ ions in the (001) plane and the layers stack in a head-to-head way along the [001] direction separated by Ba$^{2+}$ ions. We notice that O deficient perovskites are not rarely seen in Cu-based superconductors.

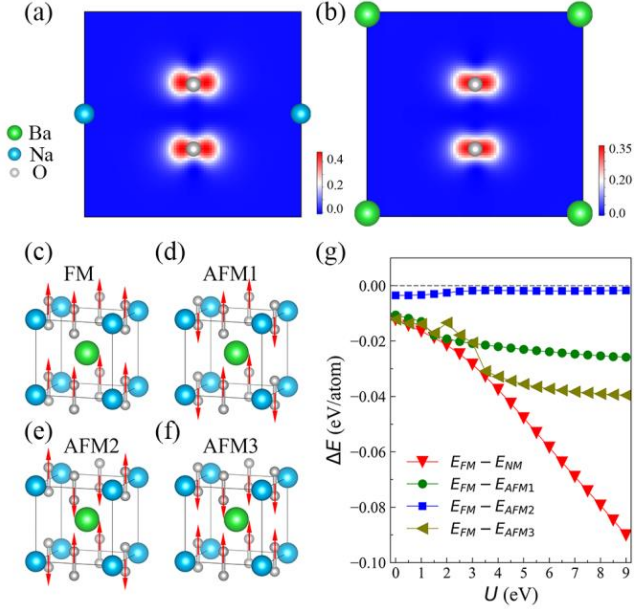

**Figure 3.** Magnetic density (electron/Å$^3$) of α-BaNaO$_4$ (a) (100) plane and (b) (200) plane. The magnetic density $\rho_{mag}$ is defined as $\rho_{mag}=\rho_{up}-\rho_{down}$, where $\rho_{up}$ is the charge density of spin up electrons, $\rho_{down}$ the charge density of spin down electrons. (c)-(f) Four possible collinear magnetic configurations. (g) Energy differences between the FM and AFM1, AFM2, AFM3, NM configuration as a function of *U*, respectively.

Then convex hull analyses of formation enthalpy were adopted to check the stability of these compounds by comparing with the known ones in the Ba-Na-O ternary system. Figure 2(g) is the calculated phase diagram. Only α-BaNaO$_4$ is stable in formation energy. Others lie above the convex hull, suggesting they are unstable or meta-stable. We further explore the behavior of α-BaNaO$_4$ in terms of lattice-dynamical, mechanical and thermal stability. As there are no any negative frequencies in the calculated phonon spectrum, it is lattice-dynamically stable, see Figure S1 deposited in the supporting information (SI). For a stable compound, its elastic stiffness constants should satisfy the Born stability criteria.[30] The calculated stiffness values of α-BaNaO$_4$ are found to well guarantee the demand of mechanical stability (see details at the SI). Furthermore, the structural stability at 500 K was examined through AIMD. During the simulation process (0~9 ps), the structure is well preserved in the framework of the Nosé-Hoover thermostat ensemble, see Figure S2 at the SI. Hence, the structure of α-BaNaO$_4$ is at least stable up to 500 K.

As well known, the prediction of the ground state of molecular magnetism in some compounds, in particular, in electron strong-correlated compounds, often fails if using LDA and GGA as exchange correlation potentials. Using GGA+*U* can correctly reproduce the magnetic ground state if *U* is properly adopted.[10,17,18,21,31,32] Here *U* is the electron correlation energy. RbO$_2$ is such a compound. Only when *U* is properly taken account for, can its electronic and the magnetic structures be calculated to be in good agreement with the experimental results. We follow this route and see whether this treatment can reproduce the right ground state. Our result shows that as *U*<4.5 eV the ground state is FM half-metal while as *U*≥4.5 eV we got the right AFM insulating state (Figure S3 at the SI) without setting any constraint on symmetry as suggested by *Ref*. 17. Therefore, using GGA+*U* is an effective way to determine the magnetic ground state in O$_2$ dimer systems. Then, for α-BaNaO$_4$, we need to know: 1) whether a moment resides in each O$_2$ dimer; and 2) how these moments order in the ground state? 3) what is its ordering temperature?

To this end, we did spin polarized calculations for α-BaNaO$_4$ with a PBE functional, and found that α-BaNaO$_4$ is a half-metal, see Figure S4(a) at the SI. The magnetic density, the difference between spin-up and spin-down electrons, indicates that magnetic moments of α-BaNaO$_4$ originate from the O$_2$ dimers, see Figure 3(a-b). Then the energies for non-magnetic (NM), FM, and other possible three AFM configurations, see Figure 3(c-f), are calculated against a series of *U* values. Figure 3(g) indicates that the FM configuration is the most favored state over all others including the NM state from *U*=0 to *U*=9 eV. These results answer the question 1) and 2). Hence, α-BaNaO$_4$ is a ferromagnet in its ground state.

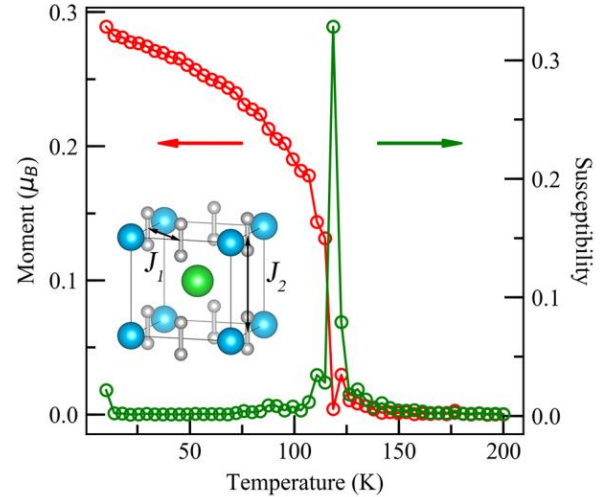

**Figure 4**. Temperature-dependent average magnetic moment and magnetic susceptibility based on Monte Carlo simulation. The inset shows indication of exchange interaction for the nearest-neighbor $J_1$ and next-nearest-neighbor $J_2$.

To answer the question 3), we resort to Monte Carlo simulations based on the classical Heisenberg model. The Hamiltonian is written as $H = -\sum_{ij} J_1 M_i M_j - \sum_{kl} J_2 M_k M_l -$

$A \sum_i (s^z)^2$, where $J_1$ and $J_2$ are the first and second nearest-neighbor exchange coupling constants, see the inset of Figure 4, and $A$ is the anisotropy energy parameter. $J_1$ and $J_1$ are extracted from Figure 3(g) by using different energies of magnetic ordering. $A$ is extracted from the calculations on varying spin orientations (see details at the SI). The obtained $J_1$, $J_2$, and $A$ are 36.9 meV, 15.9 meV, and 0.243 meV, respectively. Positive $J_1$ and $J_2$ indicate the coupling is FM. Figure 4 shows the variation of magnetic moment and magnetic susceptibility with respect to temperature. The magnetic moment decreases with increasing temperature and reaches zero at 120 K. A peak appearing at this temperature in the susceptibility indicates that $T_c$ is about 120 K.

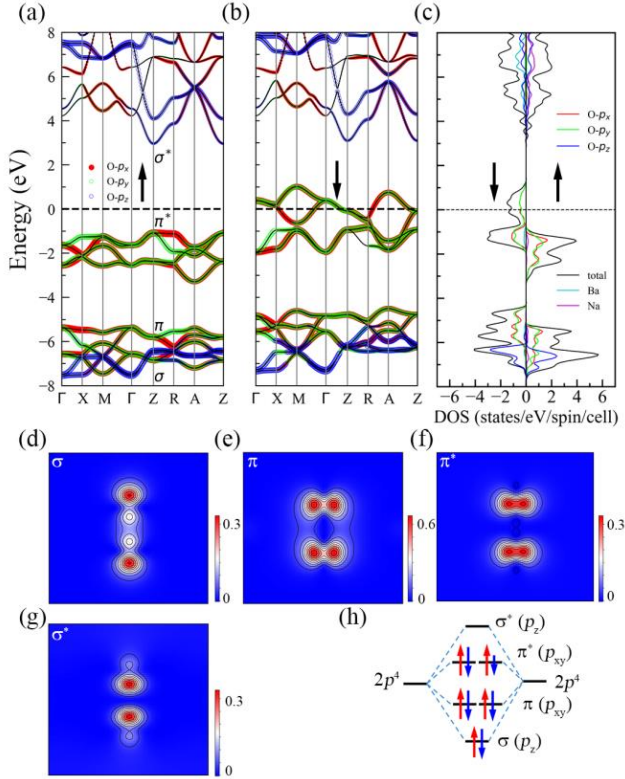

**Figure 5**. Spin and orbital resolved band structure of $\alpha$-BaNaO$_4$ for (a) spin-up subband, and (b) spin-down subband with a HSE06 hybrid functional, where the Fermi level was set to be zero. Red circles indicate O-$p_x$ orbital, green O-$p_y$, and blue O-$p_z$ orbitals. The contributions of the orbitals are proportional to the size of the circle. (c) Orbital resolved density of states contributing from Ba, Na and O atoms in $\alpha$-BaNaO$_4$. (d)-(g) charge density of (100) plane for the bands labeled in the (b). (h) Schematic representation of hybridization of O$_2$ dimer. Note here that the extra 1.5 electrons are donated from the Ba and Na atoms.

Like 3$d$ and 4$f$ metals, the local moment carried by O$_2$ dimer is due to the unpaired electrons in each $\pi^*$ orbital. For $\alpha$-BaNaO$_4$, 0.5 electron is expected to be absent in its two $\pi^*$ orbitals for each O$_2$ dimer. In this case, the spins are very likely to polarize. Figure 5(a-b) show the spin resolved band structures calculated with a HSE06 hybrid functional. In the spin-up subband, a band gap ~4 eV, separates the valence band consisting of fully occupied $\pi^*$ orbitals and the fully empty conduction band of $\sigma^*$. On the contrary, in the spin-down subband, the energy dispersions of unpaired electron in the $\pi^*$ orbitals cross the Fermi energy, resulting in a metallic feature. This is a typical half-metal. The partial density of states (Figure 5(c)) confirms that the states near the Fermi level dominantly come from the O-2$p$ orbitals. The states of the Na and Ba lie far away from the Fermi level, indicating there is nearly no orbital hybridization between O and Na/Ba. $\alpha$-BaNaO$_4$ is essentially an ionic compound.

More specifically, the Bader charge analysis indicates that Ba loses 1.58 $e^-$, Na 0.87 $e^-$, and O$_2$ dimer gains 1.23 $e^-$ averagely. Hence, the average valence state of O$_2$ should be written as O$_2^{1.5-}$ considering the charge neutrality requirement. Figure 5(d)-(g) show the charge density for the bands labeled in Figure 5(a). Obviously, the $p_z$-$p_z$ orbitals of the O$_2$ dimer form the $\sigma$ and $\sigma^*$ bonds and side-to-side overlap of $p_x$-$p_x$ and $p_y$-$p_y$ orbitals the $\pi$ and $\pi^*$ bonds. The O$_2$ molecular feature still preserves. Figure 5(h) depicts schematically the hybridization of O$_2$ dimer. In the spin-up subband, $\pi^*$ bands are fully occupied by two electrons. It should be noted that in the spin-down subband, $\pi^*$ bands are only averagely occupied by 1.5 $e^-$/dimer. In this way, the magnetic moment of per O$_2$ dimer should be 0.5 $\mu_B$, consistent with the calculated magnetic moment 1 $\mu_B$/cell (2 dimers/cell). Such an integer moment is another character of half-metals.

Minjae *et al*. argued that the spin-orbit coupling (SOC) along with Coulomb correlation will open a gap in the metallic channel of FM configuration KO$_2$.[21] We examine this effect on band structure of $\alpha$-BaNaO$_4$ by using GGA, GGA+$U$, GGA+SOC, and GGA+$U$+SOC as the exchange corrections. Our results only found that small shifts or disappearances of degeneracy in dispersions are observed for some $K$ points, see Figure S4 at the SI. Therefore, the half metallicity of $\alpha$-BaNaO$_4$ originates from partially filled $\pi^*$ bonds and SOC is not indispensable here.

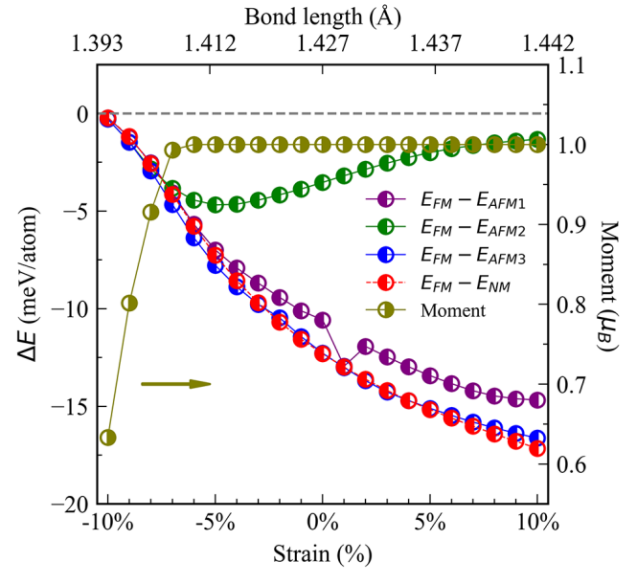

**Figure 6**. Magnetic moment of FM configuration and energy difference between AFM1, AFM2, AFM3, NM and FM vary as strains. The upper axis shows the corresponding bondlength of O$_2$ dimer at different strains.

We examine the robustness of the FM ordering under certain strains. The strain was mimicked by changing the lattice parameters, $a$ and $c$, with a ratio $\chi$, where a negative $\chi$ indicates compressive strain, and a positive χ tensile strain. The atomic positions are then optimized against the new parameters. As depicted in Figure 6, when applying compressive strains, the magnetic moment almost keeps unchanged within 6% strain then rapidly decreases. The half-metallicity will disappear because the moments are less than 1 $\mu_B$, violating the requirement of integer moment. Meanwhile, the bondlength of O-O vary from 1.427 Å to 1.393 Å. Moreover, as a compressive strain up to 20%, the $O_2$ dimers become much shorter ~1.340 Å and non-magnetic. Similar cases are observed for SrN and $SrN_2$, where SrN owns magnetic moments with a proper bondlength for the $N_2$ dimer while $SrN_2$ is non-spin polarization with a shorter bondlength.[33] So, the bondlength of the $O_2$ dimer is highly relevant to the appearance of polarized spins. After applying tensile strains within 10%, the FM keeps against three AFM and NM configurations, and the half-metallic feature retains. This means that the half-metallicity is more sensitive to compressive than to tensile strains for α-BaNaO$_4$.

The ferromagnetic ordering of α-BaNaO$_4$ could be explained by the coexistence of double exchange and direct exchange interactions. Its chemical formula can be written as $Ba^{2+}Na^+O_2^{2-}O_2^-$, i.e. mixed-valent $O_2^{2-}$ and $O_2^-$ anions coexist. Similar to $La_{1-x}A_xMnO_3$ (A=Ca, Sr, Ba), where the electron hopping from $Mn^{3+}$ to $Mn^{4+}$ through double exchange,[34] the itinerant electrons in the π* orbitals could hop from $O_2^{2-}$ to $O_2^-$ in the (001) plane mediated by $Na^+$ ions, guaranteeing the spins parallel.[20] Along the [001] direction, a positive $J_2$ could be explained as FM interaction through direct exchange interactions among the head-to-head $O_2$ dimers. All neighboring $O_2$-dimer layers are coupled ferromagnetically as expected.

α-BaNaO$_4$ is different from the known $O_2$ dimer-containing compounds in both structure and magnetic properties. Firstly, the orientation of $O_2$ dimers is a key factor. If they do not align in one direction, it is hard to form long-range magnetic ordering. That is the cases for $Rb_4O_6$ and $Cs_4O_6$, where the $O_2$ dimers do not align in an identical orientation, leading to frustrated magnetism.[10,11] Secondly, the coupling type between $O_2$ dimers is important. Although the dimers align along one direction in $AO_2$ (A=K, Rb, Cs), they show AFM orderings because of inter-AFM coupling between adjacent planes.[21] In α-BaNaO$_4$, the adjacent planes couple ferromagnetically through direct interactions. The coupling switch are attributed to the difference stacking way of $O_2$ dimer layers: a shift (0.5a, 0.5a, 0) between adjacent planes for former compounds and a head to head along the c direction for the latter. So, the direct-exchange interaction between adjacent $O_2$ dimers along the c-axis plays a vital role in the formation of ferromagnetic ordering. To check this, we shift the middle layer in AFM $RbO_2$ with a vector (0.5a, 0.5a, 0) within the unit cell, and the resultant arrangement of $O_2$ dimers between the adjacent layers is head-to-head, see Figure S5 at the SI. In this geometry, the calculated ground state changes into FM when U=6.5 eV. Noted that in a tetragonal structure, the $O_2$ dimers layers are allowed to have these two configurations only.

At low temperatures, $AO_2$ (A=K, Rb, Cs) usually undergoes several structural transitions, where the $O_2$ dimer would slightly incline with a small angle to lift the degeneracy of π* orbitals.[4,5] To clarify the influence of tilted $O_2$ dimers on the magnetic property, $O_2$ dimers in α-BaNaO$_4$ were rotated with 30°. As suggested in *Ref*. 21, two kinds of tilted structures were considered: type [R1], the $O_2$ dimers are 30° off z axis in the xz plane, see Figure S6 (a-b) at the SI; type [R2], the $O_2$ dimers are 30° off z axis in the (110) plane, see Figure S6 (c-d) at the SI. We compared the total energies of different magnetic states. It shows the inclination of $O_2$ dimers do not influence the ground state of α-BaNaO$_4$ with U=6.5 eV or without U, see Table S2 at the SI. The FM ordering still survives.

Although other predicted structures are not energy favorable, but P4/nmm-BaNaO and $P\bar{1}$-BaNaO$_2$ are stable in terms of lattice dynamics and mechanical dynamics, see details at the SI. We also examine their properties. They do not show spin polarized behavior. P4/nmm-BaNaO is a non-magnetic metal, and $P\bar{1}$-BaNaO$_2$ semiconductor, see Figure S6 and Figure S7 at the SI. It should be noted that P4/nmm-BaNaO, whose chemical formula could be written as $Ba^{2+}Na^+O^{2-}\cdot e^-$. The electrons were confined between the empty space formed by $Na^+$ ions, see Figure S7(b) at the SI. It may be a potential electride material,[35] which deserves further investigation.

We note that $N_2$ dimers can also induce magnetism in some nitrides if there exist unpaired electrons in its π* orbitals, such as SrN.[33] Besides, $N_2$ dimers were predicted to form other nitrides with main group element and 3d metals, which can be a metal with excellent thermal and electrical conductivity for $SiN_4$[36] and a half-metal for $FeN_4$.[37] More novel magnetic and other emergent properties are expected in these compounds containing $O_2$ and $N_2$ dimers.[38] Hayyan and coworkers recently gave a very comprehensive review on the generation and implications of $O_2^-$.[39] This will be helpful in synthesizing the molecule oxygen containing compounds, including these ternary ones.

■ CONCLUSION

In summary, by first principles calculations, we show that α-BaNaO$_4$ is a half-metal. It is energetically favorable, lattice-dynamically, mechanically and thermally-dynamically stable. It possesses robust FM ordering with a $T_c$ of 120 K. The partially occupied π* orbitals in the metallic channel and fully occupied π* orbitals in the semiconducting channel is the origin of half-metallicity. Unlike previously reported $AO_2$ (A=K, Rb, Cs) with AFM orderings between (001) layers through indirect exchange interactions, the $O_2$ dimer layers in α-BaNaO$_4$ coupled ferromagnetically each other through direct interactions with head-to-head alignment along the [001] direction. Our work provides new insights in exploring the FM and half-metallicity in $O_2$-dimer-containing compounds.

**ASSOCIATED CONTENT**

**Supporting Information**. Lattice parameters; elastic constants; details of estimation coupling constant; phonon spectra and band structures of BaNaO$_x$; molecular dynamics simulations of BaNaO$_4$; more band structure results of BaNaO$_4$; shifted configuration of $RbO_2$; structures and formation enthalpies of related compounds.

**AUTHOR INFORMATION**

**Corresponding Author**

jgguo@iphy.ac.cn
chenx29@iphy.ac.cn

**Notes**

The authors declare no competing financial interests.

## ACKNOWLEDGMENT

This work is financially supported by the MoST-Strategic International Cooperation in Science, Technology and Innovation Key Program (2018YFE0202601); National Natural Science Foundation of China under Grants No. 51532010, 51772322 and 51922105; the National Key Research and Development Program of China (2016YFA0300600, 2017YFA0304700); and the Key Research Program of Frontier Sciences, CAS, Grant No. QYZDJ-SSW-SLH013.

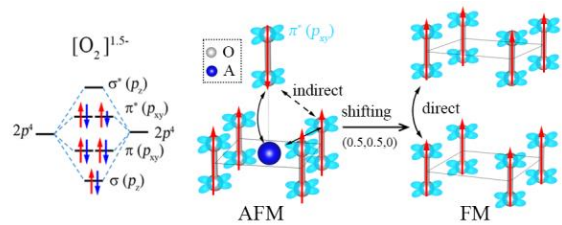